\documentclass[journal,10pt]{IEEEtran}
\usepackage{amssymb}
\usepackage{times,amsmath,epsfig,algorithmic,algorithm,color}

\ifCLASSINFOpdf
\else
\fi
\begin{document}
%
% paper title
% can use linebreaks \\ within to get better formatting as desired
% \title{Multiuser Decision Feedback with Constrains Processing for MIMO Systems. }
\title{\huge  Study of Knowledge-Aided Iterative Detection and Decoding for Multiuser MIMO Systems}
%
%
% author names and IEEE memberships
% note positions of commas and nonbreaking spaces ( ~ ) LaTeX will not break
% a structure at a ~ so this keeps an author's name from being broken across
% two lines.
% use \thanks{} to gain access to the first footnote area
% a separate \thanks must be used for each paragraph as LaTeX2e's \thanks
% was not built to handle multiple paragraphs
%

\author{Peng~Li,~\IEEEmembership{Member,~IEEE}
 Rodrigo~C.~de~Lamare,~\IEEEmembership{Senior Member,~IEEE and}
 Jingjing~Liu,~\IEEEmembership{Member,~IEEE}
        % <-this % stops a space

\thanks{%Copyright (c) 2013 IEEE. Personal use of this material is permitted. However,
%permission to use this material for any other purposes must be obtained from the IEEE by sending a request to pubs-permissions@ieee.org.
Peng~Li is with Nanjing University of Information Science and
Technology, Nanjing, China. e-mail: peng.li@nuist.edu.cn.
Rodrigo~C.~de.~Lamare is with the Centre for Telecommunications
Research (CETUC), Pontifical Catholic University of Rio de Janeiro
(PUC-Rio), Brazil and with the Communication Research Group, University of York, UK. e-mail:delamare@cetuc.puc-rio.br, rcdl500@ohm.york.ac.uk. }% <-this % stops a space
}

\maketitle

\begin{abstract}

In this work, we consider the problem of reduced latency of
low-density parity-check (LDPC) codes with iterative detection and
decoding (IDD) receiver in multiuser multiple-antenna systems. The
proposed knowledge-aided IDD (KA-IDD) system employs a minimum
mean-square error detector with refined iterative processing and a
reweighted belief propagation (BP) decoding algorithm. We present
reweighted BP decoding algorithms, which exploit the knowledge of
short cycles in the graph structure and reweighting factors derived
from the expansion of hypergraphs. Simulation results show that the
proposed KA-IDD scheme and algorithms outperform prior art and
require a reduced number of decoding iterations.

\end{abstract}


\IEEEpeerreviewmaketitle

\section{Introduction}

The fifth generation (5G) of wireless systems will demand higher
capacity, lower latency and an improved user experience \cite{5g}.
Spatially multiplexed multiuser multiple-input and multiple-output
(MIMO) systems can support several independent data streams,
resulting in a significant increase of the system throughput
\cite{Telatar}. In recent years, massive MIMO \cite{Marzetta},
\cite{NLM13}, \cite{mmimo}, \cite{wence}, \cite{mser} has been
advocated as one of the key technologies to address the capacity
requirements of 5G wireless communications. In this context, a great
deal of effort has been made in the development of detection
algorithms and their integration with channel decoding techniques
\cite{HB03}, \cite{Brink2}, \cite{Hou}, \cite{Lee}, \cite{Wu},
\cite{delamare_spadf}, \cite{Choi}, \cite{Peng}, \cite{Peng2},
\cite{Peng3}. With the adoption of modern iteratively decodable
codes such as Turbo and low-density parity-check (LDPC) codes, MIMO
systems with iterative detection and decoding (IDD) have been shown
to approach the performance of an interference free scenario.

A multiuser MIMO-IDD system is comprised of a soft-input soft-output
(SISO) MIMO detector and an efficient SISO decoder with low delay.
Specifically, the log-likelihood-ratios (LLRs) associated with the
encoded bits are updated between the two components, the information
exchange of detection and decoding is then repeated in an iterative
manner until the maximum number of iterations is reached. However,
there are many open problems for IDD schemes. These include
detection/decoding delay, which depends on the number of inner and
outer decoding iterations and performance degradation for codes with
short block lengths \cite{Brink2} \cite{Hou}.

Capacity achieving LDPC codes \cite{Gallager}, \cite{RyanLin},
\cite{bfpeg}, \cite{dopeg}, \cite{memd} are a class of linear block
codes with simple encoding and efficient decoding algorithms. The
standard belief propagation (BP) algorithm is well-known and has
been widely employed in LDPC-based IDD schemes for MIMO systems
\cite{RyanLin,Brink2,Wu,Ding,shao}. However, with the existence of
cycles in the graph structure, the standard BP has a shortcoming: at
low-to-moderate signal-to-noise ratios (SNR), a large number of
inner iterations may be required for convergence to a codeword,
which causes undesired delay and deteriorates the decoding
performance. In order to address this problem, a set of reweighting
factors have been introduced in \cite{Wainright2}, where the problem
of finding the fixed points of the BP algorithm was shown to be
equivalent to solving a variational problem. More recently,
Wymeersch et al. \cite{Wymeersch2} upgraded the reweighted BP
algorithm from pairwise graphs to hypergraphs and reduced the set of
reweighted parameters to a constant, whereas Liu and de Lamare
considered the use of two possible values in \cite{vfap}.

In this work, we present a knowledge-aided IDD (KA-IDD) scheme and
decoding algorithms for multiuser MIMO systems with reduced latency.
The proposed KA-IDD scheme and BP algorithms are inspired by the
reweighted BP decoding algorithms in \cite{Wymeersch2,vfap}, which
exploit the graphical distributions of the Tanner graph, iterative
processing and weight optimization. The proposed KA-IDD scheme
consists of a minimum mean-square error (MMSE) detector with soft
interference cancelation, refined iterative processing and a
reweighted BP decoding algorithm. We also present reweighted
knowledge-aided BP decoding algorithms: the first one is called
cycles knowledge-aided reweighted BP (CKAR-BP) algorithm, which
exploits the cycle distribution of the Tanner graph, whereas the
second is termed expansion knowledge-aided reweighted BP (EKAR-BP)
algorithm, which expands the original graph into a number of
subgraphs and locally optimizes the reweighting parameters. The
proposed KA-IDD scheme and decoding algorithms can considerably
improve the performance of existing schemes.

%{ The contributions of
%this paper can be summarized as follows:}
%
%\begin{itemize}
%\item We present a knowledge-aided IDD scheme which exploits properties of the LDPC codes to obtain improved performance with reduced processing delay.
%\item We present the CKAR-BP and EKAR-BP decoding algorithms that are incorporated into the KA-IDD scheme.
%\item We study the performance of the KA-IDD scheme and algorithms in multiuser {and massive} MIMO scenarios.
%\end{itemize}

%\begin{itemize}
%\item We first present a novel KA decoding algorithm termed cycles knowledge-aided reweighted BP (CKAR-BP) algorithm.The second KA decoding techniques is called expansion knowledge-aided reweighted BP (EKAR-BP) algorithm.
%
%\begin{itemize}
%   \item The first CKAR-BP decoder takes advantage of the cycle distribution of the Tanner graph.
%   \item The second EKAR-BP decoder expands the original graph into a number of subgraphs then locally optimizes the reweighting parameters.
% \end{itemize}
%
%
%\item Incorporated with a SISO PIC detector, both CKAR-BP and EKAR-BP algorithms are shown to outperform the standard BP and the uniformly reweighted BP (URW-BP) \cite{Wymeersch2} algorithms when performing IDD for multiuser MIMO systems.
%
%\item The proposed two decoding algorithms are incorporated into IDD schemes consists of a SISO parallel interference cancellation (PIC) scheme with a linear minimum mean-square error (MMSE) detector and their input/output extrinsic information transfer (EXIT) chart analysis is given.
%\end{itemize}

The organization of this paper is as follows: Section
\uppercase\expandafter{\romannumeral 2} introduces the system model. In
Section \uppercase\expandafter{\romannumeral 3}, the proposed
EKAR-BP and CKAR-BP algorithms are explained in detail. Section
\uppercase\expandafter{\romannumeral 4} shows the simulation results
along with discussions. Finally, Section
\uppercase\expandafter{\romannumeral 5} concludes the paper.

\section{System Model}

Let us consider the uplink of a spatially multiplexing multiuser
MIMO system with $K$ simultaneous single-antenna users and $N_R$
receive antennas ($N_R \geq K$) transmitting data over flat fading
channels. At each time instant $i$, the $K$ users transmit $K$
symbols which are organized into a $K \times 1$ vector ${\boldsymbol
s}[i] = \big[ s_1[i], ~s_2[i], ~ \ldots, ~s_k[i],~ \ldots,~ s_{K}[i]
\big]^T$ and each entry is taken from a constellation
$\mathcal{A} = \{ a_1,~a_2,~\ldots,~a_C \}$, where $(\cdot)^T$
denotes transpose and $\mathcal C$ denotes the number of
constellation points. For a given block, the symbol vector for
each user $\boldsymbol{s}_k$ is obtained by mapping it into {the
vector $\boldsymbol{x}_k = {[x_{k,1}},..., x_{k,j},..., x_{k,J}]$
with the coded bits}. The received data vector $\boldsymbol {r}[i]
\in \mathbb{C}^{N_R \times 1}$ at time instant $i$ is given by
\begin{equation}\label{1}
\boldsymbol {r}[i] = \boldsymbol{C}\boldsymbol{x}[i] +
\boldsymbol{n}[i] = \sum_{k=1}^{K}{\boldsymbol c}_k x_k[i] + {\boldsymbol n}[i],
\end{equation}
where $\boldsymbol{C} \in \mathbb{C}^{N_R \times K}$ is the channel
matrix with its {$k$th} column $\boldsymbol c_k[i] \in
\mathbb{C}^{N_R \times 1}$ representing the complex channel
coefficients, $\boldsymbol{x}[i] \in \mathbb{C}^{K \times 1}$ is the
encoded data vector with zero mean and $E\big[ {\boldsymbol x}[i]
{\boldsymbol x}^H[i]\big] = \sigma_s^2 {\boldsymbol I}$, where
$\sigma_x^2$ is the signal power, $E[ \cdot]$ stands for expected
value, $(\cdot)^H$ denotes the Hermitian operator and ${\boldsymbol
I}$ is the identity matrix. The symbol $x_k[i]$ is the encoded
transmitted bit for the $k$th user, ${\boldsymbol n}[i] \in
\mathbb{C}^{N_R \times 1}$ is complex Gaussian noise vector with
$E\big[ {\boldsymbol n}[i] {\boldsymbol n}^H[i] \big] = \sigma_n^2
{\boldsymbol I}$ with variance $\sigma_n^2$. The model in (\ref{1})
is used to represent the transmission of data symbols that are then
organized in blocks.

\section{Knowledge-Aided IDD Schemes}

In a parallel interference cancellation (PIC) based MMSE IDD
receiver, the estimates of the transmitted symbols are updated based
on the \textit{a priori} LLRs obtained from the channel decoder.
These soft symbol estimates are retrieved from the received vector
to perform interference cancellation. An MMSE filter \cite{jidf},
\cite{jiomimo} is introduced to equalize the remaining noise plus
interference term and the individual \textit{a posteriori} LLRs of
the constituent bits are obtained at the output of the filter
\cite{Wang}. According to this model in \cite{Wang}, a PIC detector
cancels the interference ($q \neq k$) with
\begin{equation} \label{2}
\centering {{\hat{\boldsymbol{r}}}_k=\boldsymbol {r}-\sum_{q \neq k}
\boldsymbol {c}_q\hat{y}_q=\boldsymbol {c}_k x_k+
{\tilde{\boldsymbol{n}}},~~~\forall k},
\end{equation}
where the co-channel interferences are estimated according to $\hat{y}_q = E[y_q] =
\sum_{a \in \mathcal{A}}P[y_q = a]a,$ where the vector $\boldsymbol {c}_k$ is the $k$th column of $\boldsymbol{C}$ and $P[y_q = a]$ corresponds to the \textit{a priori} probability of the symbol $a$ on the constellation map. Term $ {\tilde{\boldsymbol{n}}}$ is the noise-plus-remaining-interference vector to be equalized by a linear MMSE estimator as
\begin{equation} \label{3}
\centering \hat{y}_k =
{{\tilde{\boldsymbol{w}}}_k^H{\hat{\boldsymbol{r}}}_k={\tilde{\boldsymbol{w}}}_k^H\boldsymbol
{c}_k x_k+{\tilde{\boldsymbol{w}}}_k^H {\tilde{\boldsymbol{n}}}}.
\end{equation}
%where the MMSE receive filter is given by
%\begin{equation}
%{\boldsymbol{\tilde{w}}_k^H=E_x
%\boldsymbol{{c}}_k^H\big{(}\boldsymbol{C}\tilde{\boldsymbol{\Upsilon}_k}\boldsymbol{C}^H+
%\sigma_n^2 \boldsymbol{I}_{N_R}\big{)}},
%\end{equation}
%where $E_x$ is the transmitted symbol energy and
%$\tilde{\boldsymbol{\Upsilon}}_k \in \mathbb{C}^{K \times N_R}$ is a
%diagonal matrix whose entries are the variances of the estimation
%errors.

\begin{figure}[htb]
\begin{minipage}[h]{1\linewidth}
  \centering
  \centerline{\epsfig{figure=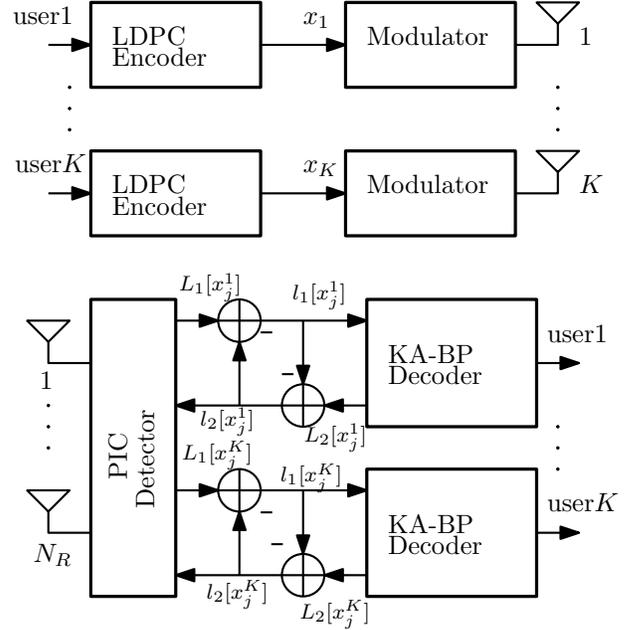,scale=1}} \vspace{-1em}
  \caption{ {Block diagram of the KA-IDD scheme for multiuser
MIMO systems.}} \label{fig:system}
\end{minipage}
\end{figure}

In Fig. \ref{fig:system}, we set $y_k = x_k +
{n}_{\scriptsize \mbox{eff}}$ at the output of the detector, where
${n}_{\scriptsize \mbox{eff}}$ is the effective noise factor after
MMSE filtering. By assuming that the output is independent from each other \cite{Wang},the approximation of the LLR of bit $x_{k,j}$:
\begin{equation}\label{4}
L_1[x_{k,j}] \approx \log
\frac{P(x_{k,j}=+1|{y_k})}{P(x_{k,j}=-1|{y_k})} =
l_1[x_{k,j}]+l_2^p[x_{k,j}],
\end{equation}
where the last term represents the \textit{a priori} information for
the coded bits $x_{k,j}$, which is obtained by the LDPC decoder. The
first term $l_1$ denotes the \textit{extrinsic} information which is
obtained by $\boldsymbol {r}[i]$ and \textit{a priori}
$l_2^p$.

The latency caused by the IDD scheme is usually due to the required inner and outer iterations involving the exchange of LLRs. The proposed KA-IDD scheme aims to reduce the number of iterations and minimizing this latency of obtaining $l_2^p[x_{k,j}]$ from the LDPC decoder.

%For the detector, by dropping the stream index $k$, the
%coded bit \textit{extrinsic} LLR is obtained as:
%\begin{equation}\label{extrif}
%l_1[x_{j}]=\log
%\frac{\sum_{a_c{\in}\mathcal{A}_{j}^+}P(y|s=a_c)\exp(L_a(a_c))}{\sum_{a_c{\in}\mathcal{A}_{j}^-}P(y|s=a_c)\exp(L_a(a_c))}
%\end{equation}
%where $\mathcal{A}_{j}^+$ and $\mathcal{A}_{j}^-$ denotes the
%subsets of constellation $\mathcal{A}$ and the bit $x_{j}$ takes the
%values 1 and 0, respectively. The value $L_a(a_c)$ denotes the
%\textit{a priori} symbol probability for symbol $a_c$ and
%\begin{equation}
% P(y | s = a_c) = \frac{1}{\pi\sigma_{\scriptsize
%\mbox{eff}}^2}\exp({\frac{-|y-s|^2}{\sigma_{\scriptsize
%\mbox{eff}}^2}})
%\end{equation}
%For the KA-IDD scheme, the computed $l_1[x_{j}]$ is fed to the LDPC
%decoder as the \textit{a priori} information and the decoder
%calculates the \textit{a posteriori} LLR of each code bit.

\section{Knowledge-Aided Decoding Algorithms}

The convergence behaviour of the BP algorithm is considered in the development of the proposed CKAR-BP and EKAR-BP algorithms. %The strategy relies on reweighting part of the hypergraph while also taking the effect of short cycles into account such that the decoder can
%generate more accurate marginal distributions.
Both algorithms relies on the techniques of reweighting part of the
hypergraph, the impact of short cycles is also considered such that
the BP decoder may calculate more accurate marginal distributions.
In \cite{Wainright2}, the reweighting strategy was employed in the
tree-reweighted BP (TRW-BP) algorithm and the authors convert BP
decoding problem to a tractable convex optimization problem,
iteratively computing beliefs and factor appearance probabilities
(FAPs). Later in \cite{Wymeersch2}, with additional constraints on
FAPs, uniformly reweighted BP (URW-BP) was introduced. Compared to
TRW-BP and URW-BP, the proposed CKAR-BP and EKAR-BP algorithms
optimize the FAPs off-line by relaxing the constraints from
\cite{Wainright2} and \cite{Wymeersch2}. Furthermore, neither of
them impose extra complexity to on-line decoding. In what follows,
we present general message passing rules for reweighted BP
algorithms, then detail the proposed CKAR-BP and EKAR-BP algorithms.

\subsection{Message Passing Rules for Knowledge-Aided Decoding}

The derivation of the message passing rules of reweighted BP
algorithms can be found in \cite{Wymeersch2} with higher-order
interactions and in \cite{Wainright2} with pairwise interactions.
Let us consider a hypergraph with $M$ check nodes, $N$ variable
nodes and the reweighting vector
$\boldsymbol{\rho}=[\rho_{1},\rho_{2},\ldots,\rho_{M}]$, the message
from the $j$th variable node $s_j$ to the $i$th check node $c_i$ is
given by
\begin{equation} \label{mp1}
\centering {\Psi_{ji}=\lambda_{\mathrm{In},j} + \sum_{i'\in
\mathcal{N}(j)\backslash
i}\rho_{i'}\Lambda_{i'j}-(1-\rho_i)\Lambda_{ij}},
\end{equation}
except $c_i$, the neighboring set of check nodes of $s_j$ is $i'\in \mathcal{N}(j)\backslash i$. Because beliefs are in the form of LLRs, $\lambda_{\mathrm{In},j}$ is equal to {
$l_1[x_{j}]$} in the first decoding iteration. We use the parameter
$\Lambda_{ij}$ to denote messages sent from $c_i$ to $s_j$ in
previous decoding iteration, then for check nodes $c_i$,
$\Lambda_{mn}$ is updated as:
\begin{equation} \label{mp2}
\centering {\Lambda_{ij}=2 \mathrm {tanh}^{-1}\big{(}\prod_{j'\in
\mathcal{N}(i)\backslash j}\mathrm{tanh}
\frac{\Psi_{j'i}}{2}}\big{)},
\end{equation}
where the hyperbolic tangent function is introduced to compute an LLR from $c_i$ to $s_j$. Finally, we have the KA-IDD updated belief $b({x_j})$ given by
\begin{equation} \label{mp3}
\centering
b({x_j})=\lambda_{\mathrm{In},j}+\sum_{i\in \mathcal{N}(j)}\rho_i\Lambda_{ij}.
\end{equation}
The proposed KA-BP algorithm employs \eqref{mp1}-\eqref{mp3} to
update the information for each node. Note that $\rho_i=1, \forall
i$ corresponds to the standard BP and negligible extra complexity is
required. At the end of the decoding procedure, the soft output is
either used for deciding the value of $\hat{x}_j$ or for generating
the extrinsic information $l_2[x_j]$ for the next KA-IDD iteration.

\subsection {Cycles Knowledge-Aided Reweighted BP (CKAR-BP)}

\begin{table}[!t]
\centering \caption{\label{tab:CKAR-BP} Proposed CKAR-BP Decoding
Algorithm}
\begin{small}
\begin{tabular}{ll}
\hline %& \tabularnewline
\textbf{{Offline Stage 1: counting of short
cycles {\scriptsize \cite{Halford}}}} & \tabularnewline
% & \tabularnewline
1: Counting the number of length-$g$ cycles {$\delta_{c_i}$} &
\tabularnewline  passing through the check node $c_i, \forall i$; &
\tabularnewline \\
% & \tabularnewline
\textbf{{Offline stage 2: determination of $\boldsymbol{\rho}_{i}$ for the hypergraph}} & \tabularnewline
% & \tabularnewline
2: Determining variable FAPs for the nodes: & \tabularnewline \textbf{if}
${ \delta_{c_i}}<\mu_g$ \textbf{then} $\rho_i=1$. \textbf{otherwise}
$\rho_i=\rho_v$ where $\rho_v= \frac{2 \alpha}{
\bar{n_D}}$; %& \tabularnewline
 & \tabularnewline \\
\textbf{{Online Stage: real-time decoding }} & \tabularnewline
% & \tabularnewline
3: Iteratively updating the belief $b(x_j)$ with reweighted &
\tabularnewline message passing \eqref{mp1}--\eqref{mp3} with
optimized $\boldsymbol{\rho}=[\rho_{1},\rho_{2},\ldots,\rho_{M}]$. &
\tabularnewline Decoding stops if { $\boldsymbol{H {\hat
x}}^T=\boldsymbol{0}$ }or the maximum iteration & \tabularnewline is
reached. & \tabularnewline
\hline & \tabularnewline
\end{tabular}\end{small}
\end{table}
The distribution of short cycles in the graph has an impact on
statistical dependency among the incoming messages being exchanged
by nodes, leading to low reliability. With the knowledge of the
cycle distribution, the proposed CKAR-BP algorithm updates the
reweighting parameters in order to mitigate the effect of short
cycles. For counting short cycles, a matrix multiplication technique
\cite{Halford} which can calculate the number of cycles with girth
of $g$, $g+2$ and $g+4$, explicitly.

In the offline stage shown in Table \ref{tab:CKAR-BP}, the parameter
$\delta_{c_i}$ denotes the number of cycles passing through check
node $c_i$ which affects the convergence behaviour of the LDPC
decoding, is determined. The average number of of length-$g$ cycles
passing a check node denoted by $\mu_g$, can be used to compute the
reweighting parameters $\rho_i (i=0, 1, \ldots, {M-1})$, we adopt a
simple criterion: %\textbf{if} $\delta_{c_i}<\mu_g~ {\rm}$
%\textbf{then}$~ \rho_i=1$, \textbf{otherwise} $\rho_i=\rho_v$,
\begin{equation}
\begin{split}
&{\rm if}  \qquad ~{ \delta_{c_i}}<\mu_g~ {\rm then} \qquad~ \rho_i=1,\\
&{\rm otherwise} ~ \rho_i=\rho_v,
\end{split}
\end{equation}
where $\rho_v=2 \alpha /\bar{n}_D$, $0< \alpha <1$ and $\bar{n}_D$
denotes the average connectivity for $N$ variable nodes given by
\begin{equation} \label{AC}
\centering {
\bar{n}_D=\frac{1}{\int_{0}^{1}{\upsilon(x)}dx}=\frac{M}{N\int_{0}^{1}{\nu(x)}dx}},
\end{equation}
where $\upsilon(x)$ and $\nu(x)$ represent the distributions of the variable nodes and the check nodes, respectively. As an improvement of URW-BP \cite{Wymeersch2}, cycle counting \cite{Halford} is required and CKAR-BP needs some extra complexity. It is important to note that when decoding LDPC codes, the proposed CKAR-BP algorithm can improve the performance of BP with either uniform structures (regular codes) or non-uniform structures (irregular codes).

\subsection {Expansion Knowledge-Aided Reweighted BP (EKAR-BP)}

\begin{table}[!t]
\centering \caption{\label{tab:EKAR-BP} Proposed EKAR-BP Decoding
Algorithm}
\begin{small}
\begin{tabular}{ll}
\hline %& \tabularnewline
\textbf{{Offline Stage 1: formation of subgraphs } } &
\tabularnewline
% & \tabularnewline
1: Applying the modified PEG expansion to generate $T\ge1$ &
\tabularnewline subgraph with a hypergraph $\mathcal{G}$ and
$d_{\mathrm{max}}$; & \tabularnewline \\
% & \tabularnewline
\textbf{{Offline Stage 2: optimization of $\boldsymbol{\rho}_{t}$ for the
$t$th subgraph}} & \tabularnewline
% & \tabularnewline
2: Initializing $\boldsymbol{\rho}_{t}^{(0)}$ to a valid value;  & \tabularnewline
% & \tabularnewline
3: For each subgraph, compute the mutual information & \tabularnewline $\boldsymbol{I}_{t}=[I_{t,1},\ldots,I_{t,L_{t}}]$  and the beliefs $b(\boldsymbol{x}_t)$ by using \eqref{mp1}--\eqref{mp3}; \\
% & \tabularnewline
4: Updating $\boldsymbol{\rho}_{t}^{(r)}$ to
$\boldsymbol{\rho}_{t}^{(r+1)}$ with the conditional gradient method & \tabularnewline provide $b(\boldsymbol{x}_t)$ and $\boldsymbol{I}_{t}$; & \tabularnewline
% & \tabularnewline
5: Repeating steps 3--4 until  each subgraph $\boldsymbol{\rho}_{t}$
converges; & \tabularnewline \\
% & \tabularnewline
\textbf{Offline Stage 3: choice of $\boldsymbol{\rho}=[\rho_{1},\rho_{2},\ldots,\rho_{M}]$ for decoding}  & \tabularnewline
% & \tabularnewline
6: For all $T$ subgraphs, collecting $\boldsymbol{\rho}_{1},\ldots,
\boldsymbol{\rho}_{i}, \ldots, \boldsymbol{\rho}_{T}$.  &
\tabularnewline and choosing the one offering the best performance;
& \tabularnewline \\
% & \tabularnewline
\textbf{{Online Stage: real-time decoding }} & \tabularnewline
% & \tabularnewline
7: Iteratively updating the belief $b(x_j)$ using reweighted &
\tabularnewline message passing rules \eqref{mp1}--\eqref{mp3} with
optimized $\boldsymbol{\rho}$. & \tabularnewline Decoding stops if
$\boldsymbol{H {\hat x}^T}=\boldsymbol{0}$ or the maximum iteration
& \tabularnewline is reached. & \tabularnewline
\hline &
\tabularnewline
\end{tabular}\end{small}
\end{table}

The proposed EKAR-BP algorithm first transforms the original
hypergraph $\mathcal{G}$ into a set of subgraphs and then locally
optimizes the reweighting parameter vector $\boldsymbol{\rho}_t,
t=1,2, \dots, T$ for each subgraph. The dimension of
$\boldsymbol{\rho}_t$ is determined by the size of the subgraph. The
TRW-BP algorithm \cite{Wainright2} (corresponds to $T=1$) has a very
slow convergence for large graphs and a computational complexity of
$\mathcal{O}(M^{2}N)$. Nevertheless, the optimization of
$\boldsymbol{\rho}$ could be significantly simpler when more
subgraphs are considered. Thus, there is need for a flexible method
to transform the original hypergraph into many subgraphs. In
general, the number of subgraphs $T$ depends on a pre-defined
maximum expansion level $d_{\mathrm{max}}$, a larger
$d_{\mathrm{max}}$ usually results in a smaller $T$ but a higher
probability of short cycles within subgraphs. Inspired by \cite{Hu},
a modified progressive-edge growth (PEG) approach is applied to
achieve the hypergraph expansion. Compared to the greedy version of
PEG \cite{Hu}, the proposed PEG expansion has two main updates:

{(i) the expansion stops as soon as every member of the set of nodes $V_t$ has been visited;}

{(ii) the number of edges incident to node $s_{j}$ might be less than
its degree since some short cycles are excluded in subgraphs to
guarantee that the local girth of each subgraph $g_t$ is larger than
the global girth of the original graph $g$.}

As shown in Table. \ref{tab:EKAR-BP}, with the obtained $T$
subgraphs, we introduce the vector
$\boldsymbol{L}=[L_{1},L_{2},\ldots,L_{T}]$ where $L_{t}$ is the
number of check nodes in the $t$th subgraph. Due to the expansion,
we have $\sum_{t}L_{t}>M$ due to duplicated nodes. Similar to TRW-BP
\cite{Wainright2}, in the $t$th subgraph, the associated FAPs
$\boldsymbol{\rho}_t=[\rho_{t,1},\rho_{t,2},\ldots,\rho_{t,L_t}]$
are optimized recursively, but with higher-order interactions and
related message passing \eqref{mp1}--\eqref{mp3}. The optimization
problem is recursively solved:

%\emph{i) for all $T$ subgraphs in parallel and fixed
%$\boldsymbol{\rho}_{t}^{(r)}$, use message passing rules
%\eqref{mp1}--\eqref{mp3} to calculate the beliefs
%$b(\boldsymbol{x}_t)$ and the mutual information term
%$\boldsymbol{I}_{t}=[I_{t,1},I_{t,2},\ldots,I_{t,L_{t}}]$ provided
%with $L_{t}\le M$ check nodes in the $t$th subgraph;}
%\\

{i) the message passing rules \eqref{mp1}--\eqref{mp3} are used to
compute the mutual information
$\boldsymbol{I}_{t}=[I_{t,1},I_{t,2},\ldots,I_{t,L_{t}}]$ and the
beliefs of $b(\boldsymbol{x}_t)$ for all $T$ parallel subgraphs and
fixed $\boldsymbol{\rho}_{t}^{(r)}$.}

{ii) given $\{\boldsymbol{I}_{t}\}_{t=1}^{T}$, we use the conditional gradient
method to update $\boldsymbol{\rho}_{t}^{(r)}$ for all $T$ subgraphs in parallel,  then go back to step 1). The objective function used by the conditional gradient method is given
by}

\begin{align*}
\mathrm{minimize} & \,\,\,\,-\boldsymbol {\rho}_t^{\dagger}\boldsymbol{I}_{t}\\
\mathrm{s.t.} & \,\,\,\,\boldsymbol{\rho}_{t} \in \mathbb{T}\big{(}\mathcal{G}_{t}\big{)},
\end{align*}
where $(\cdot)^{\dagger}$ denotes transpose,
$\mathbb{T}\big{(}\mathcal{G}_{t}\big{)}$ is the set of all valid
FAPs over the subgraph $\mathcal{G}_{t}$, and $I_{t,l}$ is a mutual
information term depending on $\boldsymbol {\rho}^{(r)}_t$, the
previous value of $\boldsymbol {\rho}_t$ representing the objective
function by $f(\boldsymbol {\rho}_t)=-\boldsymbol
{\rho}_t^{\dagger}\boldsymbol{I}_{t}$, we first linearize the
objective around the current value $\boldsymbol {\rho}^{(r)}_t$:
\begin{equation}\label{flin}
f_{\mathrm{lin}}(\boldsymbol {\rho}_t)=f(\boldsymbol {\rho}^{(r)}_t) + \nabla_{\boldsymbol {\rho}_t}^{\dagger}f(\boldsymbol {\rho}^{(r)}_t) (\boldsymbol {\rho}_t-\boldsymbol {\rho}^{(r)}_t),
\end{equation}
where$\nabla_{\boldsymbol {\rho}_t}f(\boldsymbol
{\rho}^{(r)}_t)=-\boldsymbol{I}_{t}$. Then, the term
$f_{\mathrm{lin}}(\boldsymbol {\rho}_t)$ is minimized with respect to
$\boldsymbol {\rho}_t$, denoting the minimizer by
$\boldsymbol{\rho}_t^{\ast}$ and
$z^{(r+1)}_t=\max(f_{\mathrm{lin}}(\boldsymbol{\rho}_t^{\ast}),z^{(r)}_t)$,
where $z^{0}_t=-\infty$. Finally, $\boldsymbol {\rho}^{(r)}_t$ is
updated as:
\begin{equation}\label{11}
\centering{\boldsymbol{\rho}_t^{(r+1)}=\boldsymbol{\rho}_t^{(r)}+
\alpha[\boldsymbol{\rho}_t^{\ast}-\boldsymbol{\rho}_t^{(r)}]},
\end{equation}
and $\alpha$ is obtained as:
\begin{equation} \label{10}
\arg \min_{\alpha \in [0,1]}f(\boldsymbol{\rho}_t^{(r)}+\alpha[\boldsymbol{\rho}_t^{\ast}-
\boldsymbol{\rho}_t^{(r)}]).
\end{equation}
In each recursion, $f(\boldsymbol{\rho}_t^{(r)})$ is an upper bound
on the optimized objective, while $z_t^{(r+1)}$ is a lower bound. Note that the proposed EKAR-BP algorithm can be straightforward applied if LDPC codes have been designed by the PEG principle and its variations \cite{Uchoa}, but is not limited to such designs.

\section{Simulation Results}

In this section, we present the proposed KA-IDD scheme with CKAR-BP
and EKAR-BP using an LDPC-coded uplink multiuser MIMO system with
single-antenna users. The LDPC code adopted is a regular code
designed by the PEG algorithm \cite{Hu} with block length $N=1000$,
rate $R = 0.5$, girth $g=6$, and the degree distributions are $3
(\upsilon(x)=x^4)$ and $5 (\nu(x)=x^6)$, respectively. For CKAR-BP
we employ $\alpha = 0.85$. For EKAR-BP, $T=20$ subgraphs are
generated where the check nodes are allowed to be re-visited.
EKAR-BP requires around $600$ recursions to converge for this code.

\begin{figure}[!htb]
\begin{center}
\def\epsfsize#1#2{1\columnwidth}
\epsfbox{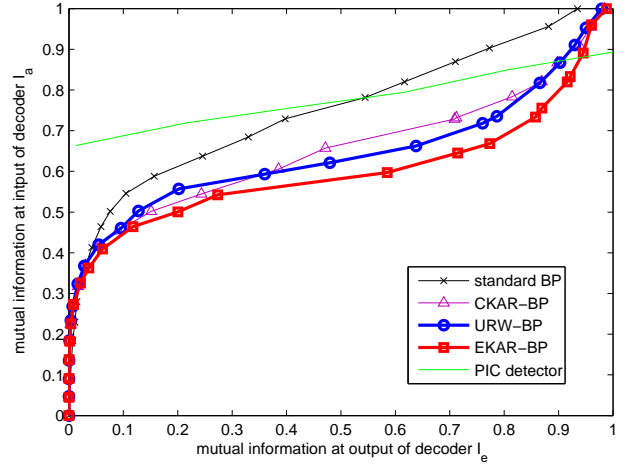}   \caption{EXIT charts of different decoders at
$E_b/N_0=4$dB. The EKAR-BP algorithm achieves a better performance
than other analyzed decoders. }\label{fig:exit}
\end{center}
\end{figure}

The iterative processing principle provides substantial gains in
each iteration. Here, we employ the extrinsic information transfer
(EXIT) chart to analyze the behavior of the constituent components
of KA-IDD scheme. Using an uncorrelated Rayleigh flat fading
channel, an EXIT chart for different decoding algorithms with the
standard BP and URW-BP algorithms are given in Fig. \ref{fig:exit}.
Even if the curve of the PIC detector does not reach the top-right
$(1,1)$ corner at the given SNR, it is obvious that the combination
of PIC detector and the proposed EKAR-BP decoding algorithm creates
the widest detection and decoding tunnel. Additionally, only the
tunnel between the PIC detector and standard BP decoding algorithm
is closed at an early stage, which indicates that performance gain
from the IDD process could be significantly diminished in this case.
To verify the result of the EXIT chart, we examine the performance
in terms of average bit-error ratio (BER).

We consider next the proposed KA-IDD scheme and decoding algorithms
in two scenarios. In the first scenario, we consider independent and
identically distributed (i.i.d) fading channel models whose
coefficients are complex Gaussian random variables with zero mean
and unit variance. In the second scenario, we consider a channel
described by
\begin{equation}
\mathbf{c}_k=\alpha_k\beta_k\mathbf{h}_k; \hspace{0.5cm}
k=1,\ldots,K, \label{eq6}
\end{equation}
where $\alpha_k$ represents the distance based path-loss between the
$k$th transmitter and the receiver, and $\beta_k$ is a log-normal
variable, representing the shadowing between the transmitter and the
receiver. The parameters $\alpha_k$ and $\beta_k$ are calculated by
%\begin{equation}
$\alpha_k=\sqrt{L_p^{(k)}}$,
%\end{equation}
and
%\begin{equation}
$\beta_k=10^{\frac{\sigma_k \mathcal{N}_k(0,1)}{10}}$,
%\end{equation}
respectively, where $L_p^{(k)}$ is the base power path loss,
$\mathcal{N}_k(0,1)$ denotes a Gaussian distribution with zero mean
and unit variance and $\sigma_k$ is the shadowing spread in dB. The
vector $\mathbf{c}_k$ in~(\ref{eq6}) is modeled as the Kronecker
channel model expressed by
\begin{equation}
\mathbf{c}_k=\mathbf{R}^{1/2}_{r_x}\mathbf{h}_{0_k},
\end{equation}
where $\mathbf{h}_{0_k}$ is the channel vector for the first
scenario and $\mathbf{R}_{r_x}$ denotes the receive correlation
matrix given by
\begin{equation}
\mathbf{R}_{r_x} = \left(
                     \begin{array}{cccc}
                       1 & \rho & \ldots & \rho^{(N_R-1)^2} \\
                       \rho & 1 & \ldots & \vdots \\
                       \vdots & \rho & \ddots & \rho \\
                       \rho^{(N_R-1)^2} & \ldots & \rho & 1 \\
                     \end{array}
                   \right).
\end{equation}
Assuming $L_p^{(k)}$, $\sigma_k$, no correlation for the $K$
transmitters with a single antenna and the correlation coefficient
$\rho = 0.8$ for all the receiver, the SNR is defined as
$10\text{log}_{_{10}}\frac{N_{t} \sigma_s^{2}}{\sigma_n^2}$, where
$\sigma_x^2$ is the variance of the received symbols and
$\sigma_n^2$ is the noise variance. The LDPC coded bits are
modulated to QPSK symbols with anti-gray coding. We used $3$ outer
detection and decoding iterations. The performance curves after $2$
outer iterations are denoted by solid lines while the curves after
$3$ outer iterations are denoted by dashed lines.

\begin{figure}[!htb]
\begin{center}
\def\epsfsize#1#2{1\columnwidth}
\epsfbox{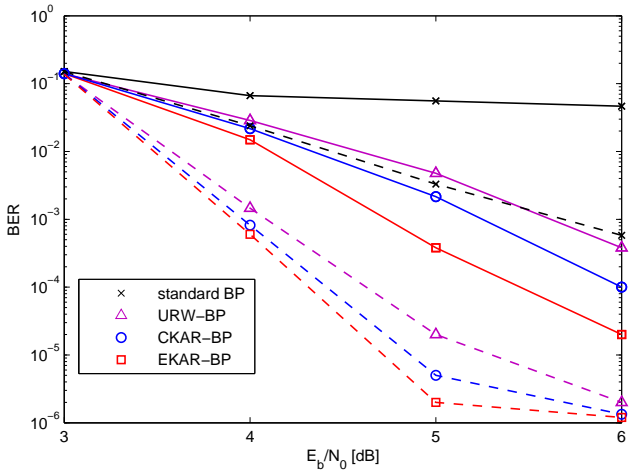} \caption{Comparison of the standard BP, URW-BP,
CKAR-BP, and EKAR-BP in terms of average BER performances for a
4-user uplink system. }\label{fig:ber}
\end{center}
\end{figure}

In the first propagation scenario shown in Fig. \ref{fig:ber}, we
employed $30$ inner decoding iterations and both CKAR-BP and EKAR-BP
decoders outperform the standard BP and URW-BP decoder in the first
detection and decoding iteration. In the third outer iteration, two
proposed decoders are still able to generate relatively good
performance when considering the low SNR range and the block length
of code.

%\begin{figure}[!htb]
%\begin{center}
%\def\epsfsize#1#2{0.65\columnwidth}
%\epsfbox{MU_MIMO_BP_URW_CKAR_EKAR.eps} \vspace{-1em} \caption{{
%Comparison of the standard BP, URW-BP, CKAR-BP, and EKAR-BP in terms
%of average BER performances for a 8-user uplink
%system.}}\label{fig:mumimo}
%\end{center}
%\end{figure}
%
%
%\begin{figure}[!htb]
%\begin{center}
%\def\epsfsize#1#2{0.65\columnwidth}
%\epsfbox{MasMIMO_BP_URW_CKAR_EKAR.eps} \vspace{-1em} \caption{{
%Comparison of the standard BP, URW-BP, CKAR-BP, and EKAR-BP in terms
%of average BER performances for a massive MIMO
%configuration.}}\label{fig:massivemimo}
%\end{center}
%\end{figure}

\begin{figure}[!htb]
\begin{center}
\def\epsfsize#1#2{1\columnwidth}
\epsfbox{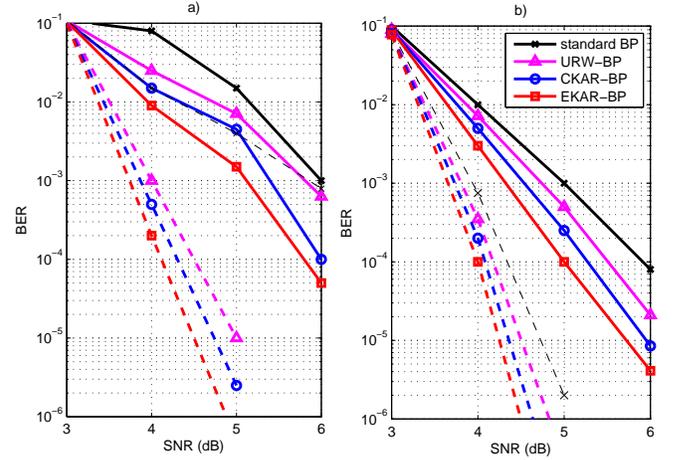} \caption{{Comparison of the standard BP, URW-BP,
CKAR-BP, and EKAR-BP in terms of average BER performances for the
uplink of a) a 8-user and b) a 8-user massive MIMO
configuration.}}\label{fig:mu&mmimo}
\end{center}
\end{figure}

In the second scenario, we have $L_{p}^{(k)}$ taken from a uniform
random variable between $0.7$ and $1$, $\tau_k = 2$ as the path loss
exponent, and the shadowing spread is $\sigma_k = 3$ dB. We employed
$20$ inner iterations and $3$ outer iterations. Fig.
\ref{fig:mu&mmimo} a) depicts a multiuser MIMO scenario with $N_R =
8$ receive antennas and $K = 8$ single-antenna users. Fig.
\ref{fig:mu&mmimo} b) demonstrate a massive multiuser MIMO case with
$N_R = 32$ receiving antennas at the base station and $K = 8$
simultaneous users. The results indicate that with a higher number
of users, the proposed algorithms also outperform the standard BP
even with a small number of outer iterations.

\section{Conclusions}

We have proposed a KA-IDD scheme for multiuser MIMO systems and two
novel KA-BP decoders, which employ reweighting strategies for
decoding regular or irregular LDPC codes. The proposed CKAR-BP and
EKAR-BP algorithms have different computational complexities in the
optimization phase and can reduce the latency caused by iterations.
The results show that the proposed KA-IDD scheme has improved
performance while using a lower number of iterations.

\end{document}